\documentclass[a4paper,amsmath,amssymb,prb,preprintnumbers,superscriptaddress,twocolumn,showpacs,longbibliography]{revtex4-1}

\pdfoutput=1 

\usepackage{graphicx}

\usepackage{dcolumn}  
\usepackage{bm}         
\usepackage{amssymb}
\usepackage{amsmath}
\usepackage{epsfig}
\usepackage{xcolor}

\usepackage{xspace}
\newcommand{\unit}[1]{\ensuremath{\,\mathrm{#1}}\xspace} 
\newcommand{\ipnm}{\ensuremath{\,\mathrm{ions/nm}}\xspace}
\newcommand{\ipsqnm}{\ensuremath{\,\mathrm{ions/nm^2}}\xspace}

\begin{document}

\title{Josephson junctions and SQUIDs created by focused helium ion beam irradiation of YBa$_2$Cu$_3$O$_7$}

\author{B.~M\"uller}
\email{benedikt.mueller@uni-tuebingen.de}
\affiliation{%
Physikalisches Institut -- Experimentalphysik II and Center for Quantum Science (CQ) in LISA$^+$,
University of T\"ubingen,
Auf der Morgenstelle 14,
72076 T\"ubingen, Germany}

\author{M.~Karrer}
\affiliation{%
Physikalisches Institut -- Experimentalphysik II and Center for Quantum Science (CQ) in LISA$^+$,
University of T\"ubingen,
Auf der Morgenstelle 14,
72076 T\"ubingen, Germany}

\author{F.~Limberger}
\affiliation{%
Physikalisches Institut -- Experimentalphysik II and Center for Quantum Science (CQ) in LISA$^+$,
University of T\"ubingen,
Auf der Morgenstelle 14,
72076 T\"ubingen, Germany}

\author{M.~Becker}
\affiliation{%
Physikalisches Institut -- Experimentalphysik II and Center for Quantum Science (CQ) in LISA$^+$,
University of T\"ubingen,
Auf der Morgenstelle 14,
72076 T\"ubingen, Germany}
\affiliation{%
NMI Natural and Medical Sciences Institute at the University of T\"ubingen,
Markwiesenstr.~55,
72770 Reutlingen, Germany}

\author{B.~Schr\"oppel}
\affiliation{%
NMI Natural and Medical Sciences Institute at the University of T\"ubingen,
Markwiesenstr.~55,
72770 Reutlingen, Germany}

\author{C.~J.~Burkhardt}
\affiliation{%
NMI Natural and Medical Sciences Institute at the University of T\"ubingen,
Markwiesenstr.~55,
72770 Reutlingen, Germany}

\author{R.~Kleiner}
\affiliation{%
Physikalisches Institut -- Experimentalphysik II and Center for Quantum Science (CQ) in LISA$^+$,
University of T\"ubingen,
Auf der Morgenstelle 14,
72076 T\"ubingen, Germany}

\author{E.~Goldobin}
\affiliation{%
Physikalisches Institut -- Experimentalphysik II and Center for Quantum Science (CQ) in LISA$^+$,
University of T\"ubingen,
Auf der Morgenstelle 14,
72076 T\"ubingen, Germany}

\author{D.~Koelle}
\affiliation{%
Physikalisches Institut -- Experimentalphysik II and Center for Quantum Science (CQ) in LISA$^+$,
University of T\"ubingen,
Auf der Morgenstelle 14,
72076 T\"ubingen, Germany}

\date{\today}

\begin{abstract} 
By scanning with a $30\, \mathrm{keV}$ focused He ion beam (He-FIB) across YBa$_2$Cu$_3$O$_7$ (YBCO) thin film microbridges, we create Josephson barriers with critical current density $j_\mathrm{c}$ adjustable by irradiation dose $D$.
The dependence $j_\mathrm{c} (D)$ yields an exponential decay.
At $4.2\, \mathrm{K}$, a transition from flux-flow to Josephson behavior occurs when $j_\mathrm{c}$ decreases below $\approx 2\, \mathrm{MA/cm^2}$.
The Josephson junctions exhibit current-voltage characteristics (IVCs) that are well described by the resistively and capacitively shunted junction model, without excess current for characteristic voltages $V_\mathrm{c} \lesssim 1\, \mathrm{mV}$.
Devices on MgO and 
LSAT substrates show non-hysteretic IVCs, while devices on SrTiO$_3$ show a small hysteresis.
For all junctions an approximate scaling $V_\mathrm{c} \propto j_\mathrm{c}^{1/2}$ is found.
He-FIB irradiation with high dose produces barriers with $j_\mathrm{c}=0$ and high resistances of $10\, \mathrm{k\Omega} \ldots 1\, \mathrm{G\Omega}$.
This provides the possibility to write highly resistive walls or areas into YBCO using a He-FIB.
Transmission electron microscopy reveals an amorphous phase within the walls, whereas for lower doses the YBCO stays crystalline.
We have also ``drawn'' superconducting quantum interference devices (SQUIDs) by using a He-FIB for definition of the SQUID hole and the junctions.
The SQUIDs show high performance, with flux noise $< 500\, \mathrm{n \Phi_0/Hz^{1/2}}$ in the thermal white noise limit for a device with $19\, \mathrm{pH}$ inductance.
\end{abstract} 

\pacs{%
85.25.CP, 
85.25.Dq, 
74.78.Na, 
74.72.-h 
74.25.F- 
74.40.De 
%
}


\maketitle

\section{Introduction}
\label{sec:Introduction}

Josephson junctions (JJs), i.e., weak links between two superconducting electrodes~\cite{Likharev79}, are key elements in superconducting electronic circuits and are used both for basic studies of superconductivity and for many applications~\cite{Barone-Paterno82, Kleiner-Buckel16}.
For conventional metallic superconductors, a mature trilayer thin film technology based on Nb electrodes, separated by insulating or normal conducting barriers has been well established for decades.
This technology offers fabrication of JJs on wafer scale with small spread of characteristic parameters, such as critical currrent density $j_0$ and normal resistance times area $\rho_\mathrm{n}$, even with lateral JJ size well below $1 \unit{\mu m}$~\cite{Bhushan91,Hagedorn06}.

For the high-transition temperature (high-$T_\mathrm{c}$) cuprate superconductors, JJ technology is much less mature.
Due to the complex nature of these materials, and in particular due to their small coherence length associated with strong sensitivity to defects on the atomic scale, a reliable trilayer JJ technology does not exist so far.
On the other hand, the peculiar properties of cuprate superconductors, such as high $T_\mathrm{c}$, large upper critical field, large energy gap and $d$-wave symmetry of the superconducting order parameter, can provide major advantages, if JJ devices and circuits can be realized with sufficient control over JJ parameters.
Promising examples are e.g.~in the field of THz generation~\cite{Welp13}, self-biased rapid single flux quantum circuits~\cite{Ortlepp06}, or magnetometry based on superconducting quantum interference devices (SQUIDs)~\cite{Koelle99,Faley17}.

Apart from intrinsic JJs in stacks of Bi$_2$Sr$_2$CaCu$_2$O$_{8+\delta}$ single crystals, used for THz generation~\cite{Welp13}, most developed cuprate JJs are based on epitaxially grown YBa$_2$Cu$_3$O$_{7-\delta}$ (YBCO) thin films with $T_\mathrm{c}\approx 90 \unit{K}$, that offers also operation with cryocoolers or liquid nitrogen, and a large variety of JJ types have been developed and their properties investigated~\cite{Gross97,Hilgenkamp02,Tafuri05,Tafuri13}.

Until today, the most reliable, simple and most frequently used high-$T_\mathrm{c}$ JJs are YBCO grain boundary (GB) JJs~\cite{Hilgenkamp02}.
They are usually fabricated by the epitaxial growth of YBCO films on (rather expensive) bi-crystal substrates from only a few materials.
The GBJJs can be placed only along the single GB line, which not only imposes topological limitations, but also limits the complexity of feasible circuits.
The more advanced biepitaxial technique allows one to fabricate so-called tilt-twist GBs~\cite{Tafuri05}.
Such GBJJs can be distributed all over the chip and one can even fabricate $0$-$\pi$ JJs~\cite{Cedergren10a}.
Still, GBJJs suffer from hardly controllable inhomogeneity along the GB line, which makes the properties of the JJs not very reproducible and causes a substantial spread in JJ parameters.
Alternative approaches to create Josephson barriers in cuprates are based on local irradiation of thin films with a high-energy focused electron beam~\cite{Tolpygo93,Pauza97,Booij97,Booij99} or on irradiation with high-energy ions (protons~\cite{Booij99}, neon~\cite{Chen04,Chen05a}, oxygen~\cite{Tinchev90,Bergeal05}) through a lithographically defined mask with a nanogap.
The local irradiation drives the material from the superconducting to the normal conducting or even insulating state.
So far, this approach was hampered by the fact that it was not possible to create ultrathin Josephson barriers that would provide JJs with high characteristic voltage $V_\mathrm{c} = j_0 \rho_\mathrm{n}$ and current-voltage characteristics (IVCs) without excess current that are well described by the resistively and  capacitively shunted junction (RCSJ) model~\cite{Stewart68,McCumber68}. This is an important prerequisite for many applications.
For reviews on various approaches to modify the properties of cuprate superconductors by local irradiation see e.g.~Refs.~[\onlinecite{Cybart14}], [\onlinecite{Lang-NSES-3}] and references therein.

With the recent development of Helium ion microscopy (HIM) \cite{HIM}, a sharply focused He ion beam with $\sim 0.5 \unit{nm}$ diameter can be used to irradiate and modify cuprate superconductors on the nanoscale.
This approach has been successfully used by Cybart and co-workers to fabricate JJs by focused Helium ion beam (He-FIB) irradiation of YBCO thin films, and they demonstrated that the barriers in such He-FIB JJs can be changed continuously from conducting to insulating by varying the irradiation dose~\cite{Cybart15}.
Moreover, the same group demonstrated already the integration of He-FIB JJs into SQUID devices~\cite{Cho15}, and the feasibility to use high-dose irradiation for nanoscale patterning (without removing material) in YBCO devices~\cite{Cho18,Cho18a}.
For a short review on this approach, also including irradiation with a focused  Ne ion beam, see Ref.~[\onlinecite{Cybart-HIM-17}].
First attempts to extend this technique to the fabrication of JJs in other cuprate materials have been reported~\cite{Gozar17}, and the creation of JJs in MgB$_2$ thin films by He-FIB irradiation has been demonstrated very recently~\cite{Kasaei18}.

Here we report on the realization of He-FIB JJs in YBCO thin films on different substrates.
We focus on the analysis of the electric transport properties at 4.2\,K of such JJs, complemented by numerical simulations based on the RCSJ model, and on the dependence of the JJ properties on irradiation dose.
We also present results on scanning transmission electron microscopy analysis of the local structural modification of the YBCO films which can be made highly resistive by He-FIB irradiation with high dose.
The latter feature has been used to fabricate SQUIDs, by combination of medium-dose irradiation to produce two JJs with high-dose irradiation to produce the SQUID loop, and we demonstrate dc SQUID operation, including low-noise performance.

\section{Device fabrication}
\label{sec:Fab-Setup}

We fabricate epitaxially grown $c$-axis oriented YBCO thin films of thickness $d$ by pulsed laser deposition (PLD) on various single crystal (001)-oriented SrTiO$_3$ (STO), MgO and (LaAlO$_3$)$_{0.3}$(Sr$_2$AlTaO$_6$)$_{0.7}$ (LSAT) substrates ($10 \times 10 \unit{mm^2}$).
The crystalline quality of the YBCO films is characterized by x-ray diffraction to determine the full width half maximum (FWHM) of the rocking curve of the YBCO (005) peak and to extract $d$ via Laue oscillations at the YBCO (001) Bragg peak.
YBCO films on STO substrates are covered in-situ by an epitaxially grown, 10 unit cells ($3.9 \unit{nm}$) thick, STO cap layer.
For details on PLD growth of our YBCO films on STO substrates, and their structural and electric transport properties see Refs.~[\onlinecite{Werner10}] and~[\onlinecite{Schwarz13}].

For electrical contacts on STO and MgO chips, we photolithographically define a resist mask, covering the central area of the chips, remove a $\sim 10 \unit{nm}$ thick surface layer (including the STO cap layer) by Ar ion milling and in-situ deposit a Au film by magnetron sputtering, followed by a lift-off process.
For electrical contacts on LSAT chips, we deposit in-situ, after the PLD process, a Au film onto the YBCO by in-situ electron beam evaporation after the PLD process.
Subsequently, we photolithographically define a resist mask, covering the outer area of the chips, and remove the Au film on the central area of the chip by Ar ion milling.

\renewcommand{\arraystretch}{1.3}
\begin{table}[b]
\caption{Properties of studied chips with YBCO thin film microbridges used for fabricating He-FIB-irradiated devices.
To calculate the effective penetration depth $\lambda_\mathrm{eff}=\lambda_\mathrm{L}^2/d$ we assume a London penetration depth $\lambda_\mathrm{L} = 250 \unit{nm}$.}
\begin{center}
\tabcolsep1.5mm
\begin{tabular}{p{2cm} c c c c c c c c c c}\hline\hline
chip 			& $T_\mathrm{c}$	& FWHM            	& $d$ 	& $w$        	& $\lambda_\mathrm{eff}$\\
(substrate)	& (\unit{K})        		& ($^\circ$)		& (\unit{nm})	& ($\unit{\mu m}$)	& ($\unit{\mu m}$)\\\hline\hline
STO-1		& 89				& 0.11      		& 50    	& 3.2 		& 1.25\\\hline
STO-2		& 89      			& 0.15      		& 29    	& 4.0 		& 2.16\\\hline
STO-3		& 91      			& 0.08     		& 46    	& 3.8 		& 1.36\\\hline
MgO-1		& 89      			& 0.17      		& 53    	& 4.1 		& 1.18\\\hline
LSAT-1		& 86      			& 0.07      		& 50  	& 4.4 		& 1.25\\\hline
\end{tabular}
\end{center}
\label{Tab:samples}
\end{table}

Afterwards, we use photolithography and Ar ion milling to prepattern 156 YBCO microbridges of width $w \approx 4 \unit{\mu m}$ (and length $\approx 200 \unit{\mu m}$) on each chip for He ion irradiation and electric transport measurements in a four-point configuration.
Table~\ref{Tab:samples} gives an overview on the five chips, with some basic properties of their YBCO microbridges, which have been used for fabricating devices via He-FIB irradiation.
A specific bridge on one of the chips is labeled by the chip name, followed by \#$n$ for bridge number $n$, e.g.~STO-1\#4 corresponds to bridge number 4 on the chip STO-1.

After prepatterning the YBCO microbridges, focused He ion beam irradiation is done in a Zeiss Orion NanoFab He/Ne ion microscope (HIM) with 30\,keV He$^+$ ions.
A beam current of $200 \unit{fA}$ is used, and the beam is focused to a nominal diameter of $0.5 \unit{nm}$.
A dwell time of $1 \unit{\mu s}$ is used to irradiate line patterns with a dwell point spacing of $0.25 \unit{nm}$, which corresponds to a single line scan dose $D_\mathrm{sl} = 5 \ipnm$.
To obtain a certain line dose $D=Nd_\mathrm{sl}$, a single linescan is repeated $N$~times.
To irradiate an area, adjacent linescans are offset by $\Delta = 0.25 \unit{nm}$.
In that case, a line dose of e.g.~$D= 100 \ipnm$ corresponds to an area dose of $D_\mathrm{A} \equiv D/\Delta= 400 \ipsqnm$ or $4\times 10^{16}\unit{ions/cm^2}$.

\begin{figure}[t]
\includegraphics[width=\columnwidth]{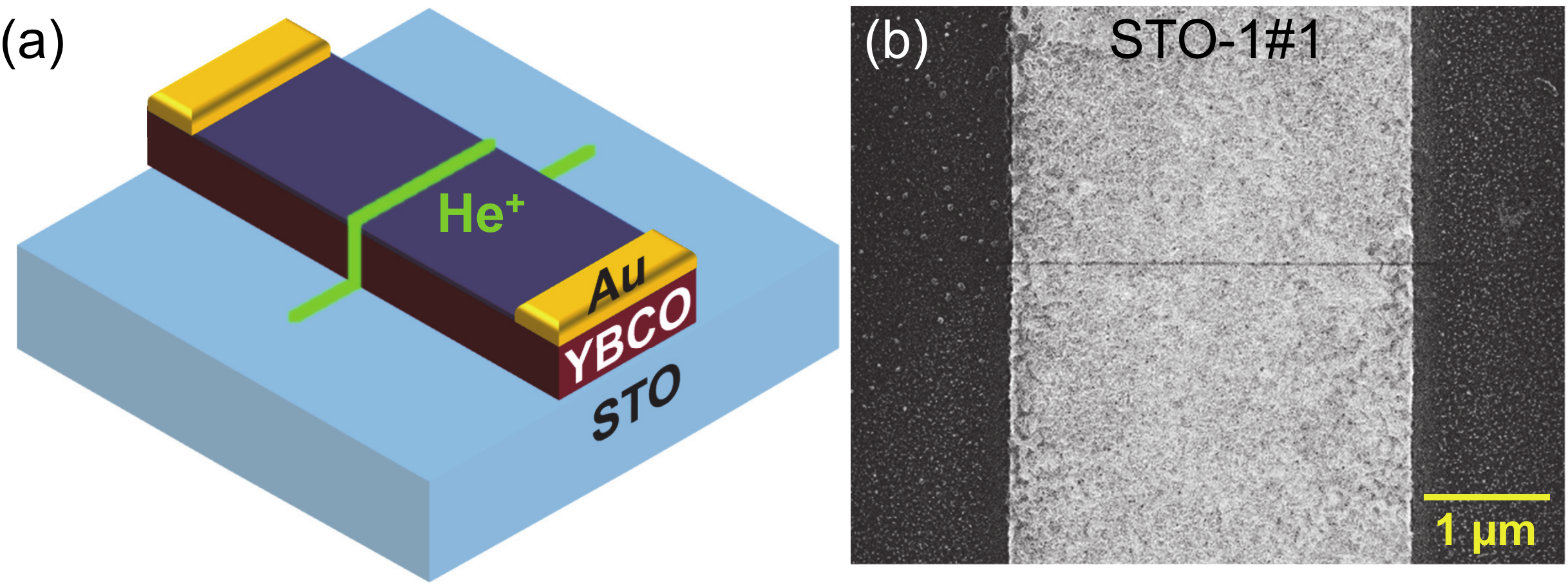}
\caption{(a) Schematic illustration of the He-FIB JJ geometry.
(b) SEM image of a JJ (visible as thin dark line) fabricated with $D=600 \ipnm$.
}
\label{Fig:sample}
\end{figure}

Figure~\ref{Fig:sample}(a) schematically illustrates the sample geometry and irradiation process for a single JJ.
A scanning electron microscopy (SEM) image of STO-1\#1 fabricated with $D= 600 \ipnm$ is shown in Fig.~\ref{Fig:sample}(b).
The irradiated linescan appears as a dark line in the SEM image due to He-FIB induced carbon deposition from residual gas inside the He ion microscope chamber.

\section{YBCO bridges with He-FIB-induced barriers and Josephson junctions}
\label{sec:bridges}

In this section, we present results obtained from devices fabricated on the chips listed in Table~\ref{Tab:samples}.

\subsection{Resistance vs.~temperature}
\label{subsec:R-vs-T}

\begin{figure}[t]
\includegraphics[width=\columnwidth]{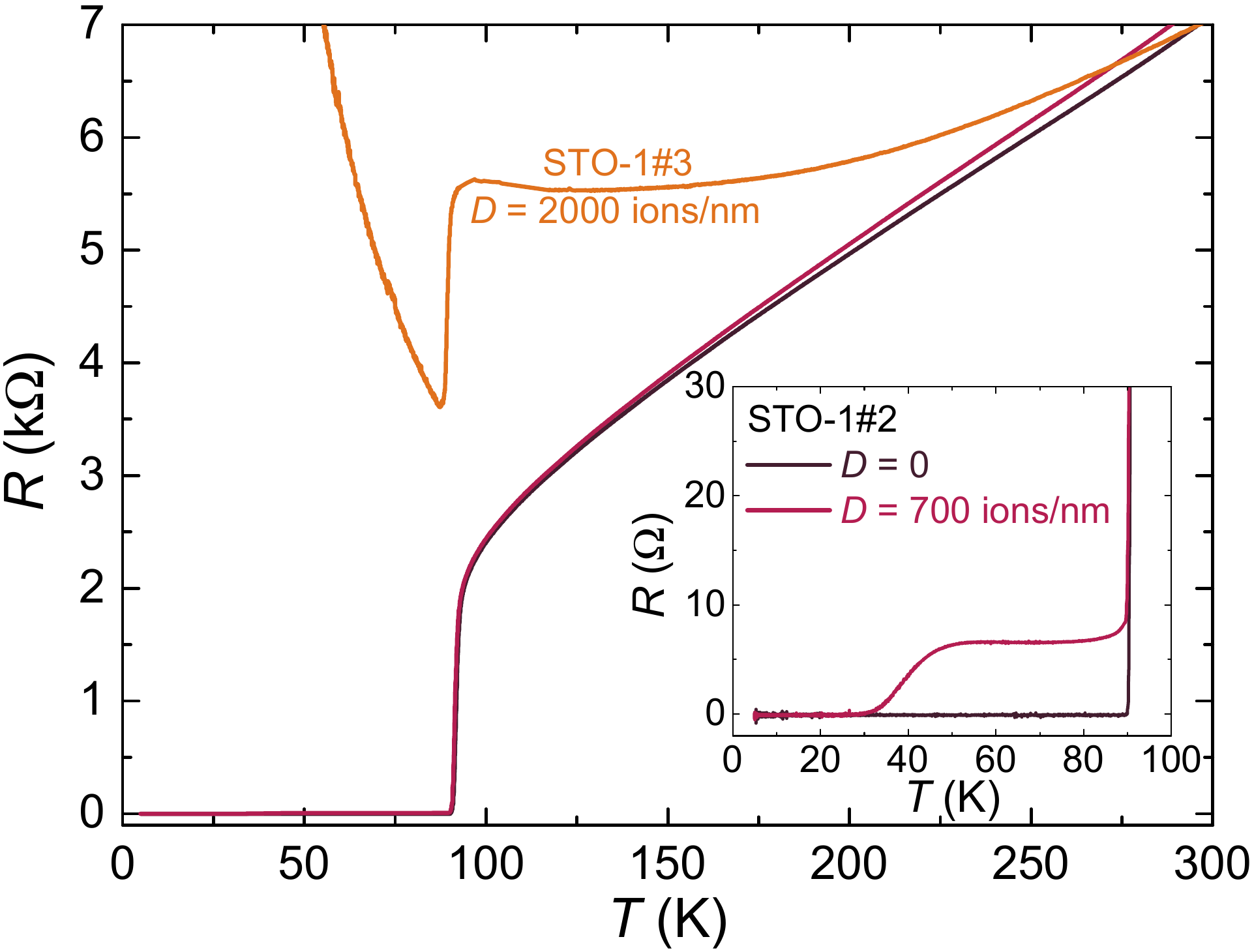}
\caption{$R (T)$ dependence of YBCO bridge STO-1\#2 before and after irradiation 
and of STO-1\#3 after irradiation.
%
Inset shows a zoom of the resistive transitions of STO-1\#2.}
\label{Fig:R(T)}
\end{figure}

Figure ~\ref{Fig:R(T)} shows measurements of the resistance $R$ (at constant bias current $I_\mathrm{b} =1 \unit{\mu A}$) vs.~temperature $T$ of two YBCO microbridges.
The $R(T)$ curve of STO-1\#2, measured before He ion irradiation, shows a decrease of the resistance by about a factor of 3 from 300 to 100\,K, with resistivity $\rho(100 \unit{K})\approx 190 \unit{\mu\Omega \, cm}$, followed by a sharp transition to $R < 1 \unit{\Omega}$ at $T_\mathrm{c} = 89 \unit{K}$.
After irradiation with $D=700 \ipnm$ (and thus producing a JJ), the $R (T)$ curve of STO-1\#2 shows an additional foot-like structure with a plateau at $R = 6.6 \unit{\Omega}$ between approximately $40 \unit{K}$ and $T_\mathrm{c}$ (see inset).
This foot structure is due to thermally activated phase slippage~\cite{Gross90a} causing a finite voltage drop across the JJ when (upon increasing $T$) the thermal energy $k_\mathrm{B} T$ approaches the Josephson coupling energy $E_\mathrm{J}=I_0 \Phi_0 / (2 \pi)$.
Here, $I_0$ is the noise-free critical current of the JJ (which decreases with increasing $T$) and $\Phi_0$ is the magnetic flux quantum.
Accordingly, the plateau reflects the situation when the measurable critical current $I_\mathrm{c}$ drops below the bias current $I_\mathrm{b}$, causing the JJ to reach its normal state resistance $R_\mathrm{n}$.
He-FIB irradiation with high dose fully suppresses $I_\mathrm{c}$.
This is shown in Fig.~\ref{Fig:R(T)} for sample STO-1\#3, which has been irradiated with $D=2000 \ipnm$.
At $T=4.2 \unit{K}$ the resistance is $> 20 \unit{k \Omega}$.

\subsection{Transmission electron microscopy analysis}
\label{subsec:TEM}

By the combination of atomic force microscopy and scanning near-field optical microscopy, it has been shown by Gozar {\it et al.}~\cite{Gozar17} that He-FIB irradiation with doses above $10^{18} \unit{ions/cm^2}$ induces amorphous tracks in La$_{1.84}$Sr$_{0.16}$CuO$_4$ thin films with substantial lateral width $\sim 500 \unit{nm}$.
In contrast to this, He-FIB produced JJs in YBCO films have been reported to show IVCs well described by the RCSJ model, indicating much less lateral damage \cite{Cybart15}.
However, no results have been reported on microstructural changes induced by a He-FIB in YBCO films so far.

\begin{figure*}[t]
\includegraphics{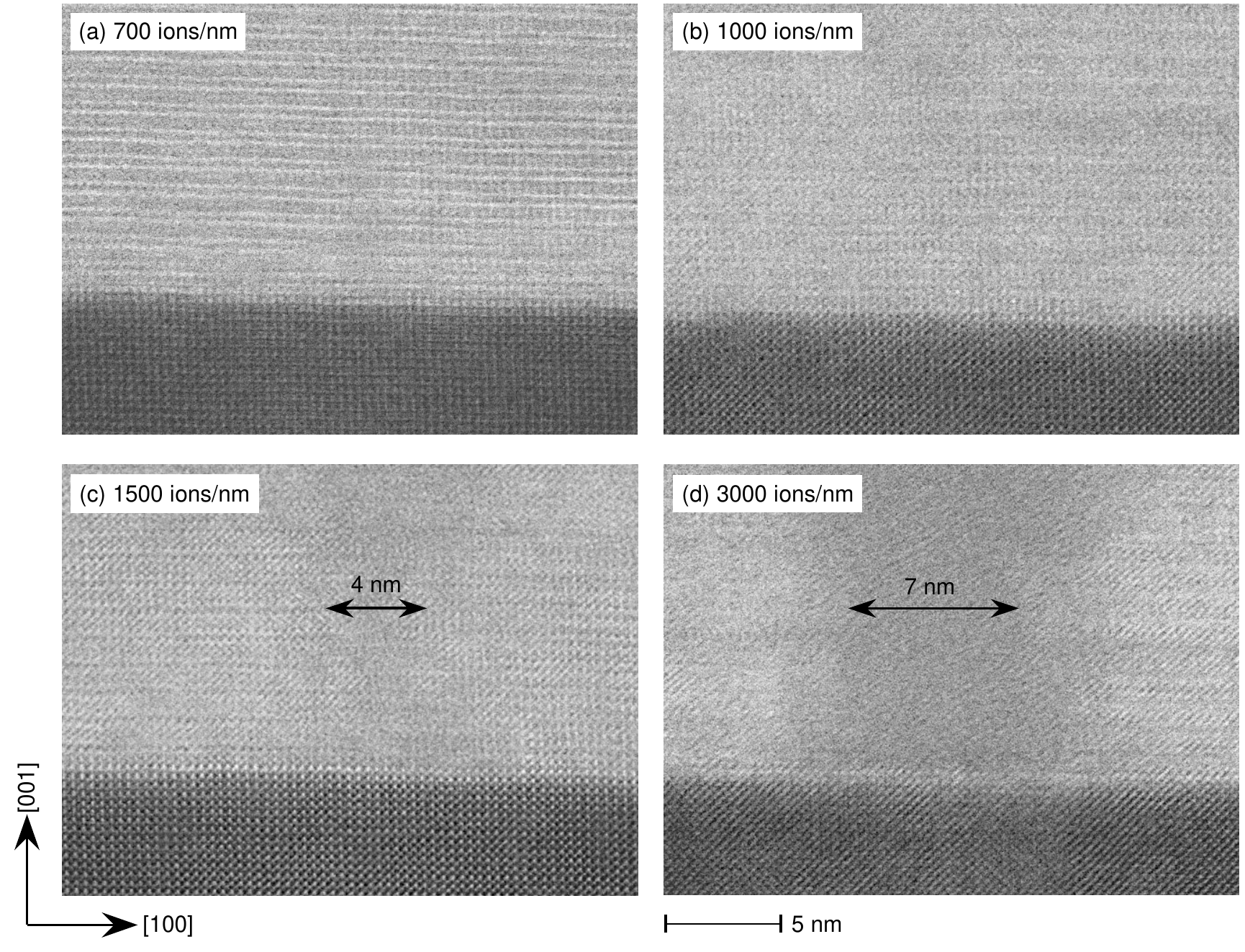}
\caption{Cross-section STEM images of YBCO (top) / STO (bottom) interface at the location of He-FIB irradiation with increasing dose $D$ from (a) to (d).
Arrows indicate widths of amorphous regions.}
\label{Fig:STEM}
\end{figure*}

To image possible structural modifications induced by He-FIB irradiation in our YBCO films, we used scanning transmission electron microscopy (STEM).
For the STEM studies, we irradiated the YBCO bridge STO-1\#4 with a series of 14 parallel lines using increasing doses from $D=50$ to $10^5 \ipnm$, with well defined spacing (200\, nm in most cases) between adjacent lines.
%
Subsequently, we prepared a cross-sectional TEM-lamella containing all irradiated lines, by in-situ lift-out using a Ga-FIB microscope together with a micromanipulator.
Figures \ref{Fig:STEM}(a) to (d) show cross-section STEM images, viewed along the [010] zone axis, of four regions of the bottom part of the YBCO film at the YBCO/STO interface that have been irradiated with $D=700$, 1000, 1500 and $3000 \ipnm$, respectively.
The areas that were irradiated with $D\ge 3000\,$ions/nm can be easily located in the STEM images due to significant changes in the microstructure of the YBCO films. As we know the exact spacing between the different areas irradiated along the lamella, we can also easily localize in the STEM images the areas that have been irradiated with lower $D$.
For $D=700$ and $1000 \ipnm$ [cf.~Fig.~\ref{Fig:STEM}(a,b)], we cannot identify any change in the structure of the irradiated sections.
For $D=1500 \ipnm$, an amorphous track of width $w_\mathrm{a} \approx 4 \unit{nm}$ appears [cf.~Fig.~\ref{Fig:STEM}(c)], increasing to $w_\mathrm{a} \approx 7 \unit{nm}$ for $D=3000 \ipnm$ [cf.~Fig.~\ref{Fig:STEM}(d)].
We note that with further increasing $D$ the amorphous track width $w_\mathrm{a}$ increases roughly linearly up to $\approx 170 \unit{nm}$ for the highest dose $D=10^5 \ipnm$ that we have investigated.

Our STEM analysis indicates that medium doses do not induce significant structural damage of the YBCO crystal lattice, which is consistent with the assumption that the He-FIB easily moves oxygen ions from the Cu-O chains to interstitial  sites~\cite{Cho18} and thereby altering the local electric transport properties of YBCO on the nanometer scale without destroying the crystal lattice as a whole.
Hence, He-FIB irradiation with medium dose seems to be a very promising approach for creating JJs in YBCO.
Moreover, we find that irradiation with larger doses of some $1000 \ipnm$ induces amorphous, and hence presumably highly resistive, regions, but still with a relatively small lateral extension of only a few nm.

\subsection{Transport characteristics of He-FIB-induced Josephson junctions}
\label{subsec:bJJs}

\renewcommand{\arraystretch}{1.3}
\begin{table*}[t]
\caption{Irradiation doses and device parameters of JJs shown in Fig.~\ref{Fig:IVC-Ic(B)}.
}
\begin{center}
\tabcolsep1.5mm
\begin{tabular}{l c c c c c c c c c c c} \hline \hline
& $D$ & $I_0$ & $R_\mathrm{n}$ & $V_\mathrm{c}$ & $C$ & $\beta_\mathrm{C}$ & $\lambda_\mathrm{J}$ & $\lambda_\mathrm{J}^\mathrm{NL}$ & $B_{\mathrm{c}1}^\mathrm{ex}$ & $B_{\mathrm{c}1}^\mathrm{LO}$ & $B_{\mathrm{c}1}^\mathrm{NL}$ \\
& ($\mathrm{ions} / \mathrm{nm}$) & ($\mathrm{\mu A}$) & ($\Omega$) & ($\mathrm{\mu V}$) & ($\mathrm{pF}$) & & ($\mathrm{\mu m}$) & ($\mathrm{\mu m}$) & ($\mathrm{mT}$) & ($\mathrm{mT}$) & ($\mathrm{mT}$) \\ \hline \hline
STO-1\#5 & 700 & 99 & 7.44 & 737 & 0.17 & 2.77 & 0.42 & 1.8 & 0.31 & 0.63 & 0.28 \\ \hline
MgO-1\#1 & 500 & 178 & 3.07 & 546 & 0.15 & 0.77 & 0.37 & 2.6 & 0.25 & 0.75 & 0.17 \\ \hline
LSAT-1\#1 & 200 & 260 & 2.69 & 699 & 0.15 & 0.83 & 0.30 & 3.5 & 0.76 & 0.88 & 0.15 \\ \hline
\end{tabular}
\end{center}
\label{Tab:devices}
\end{table*}

\begin{figure}[b]
\includegraphics[width=\columnwidth]{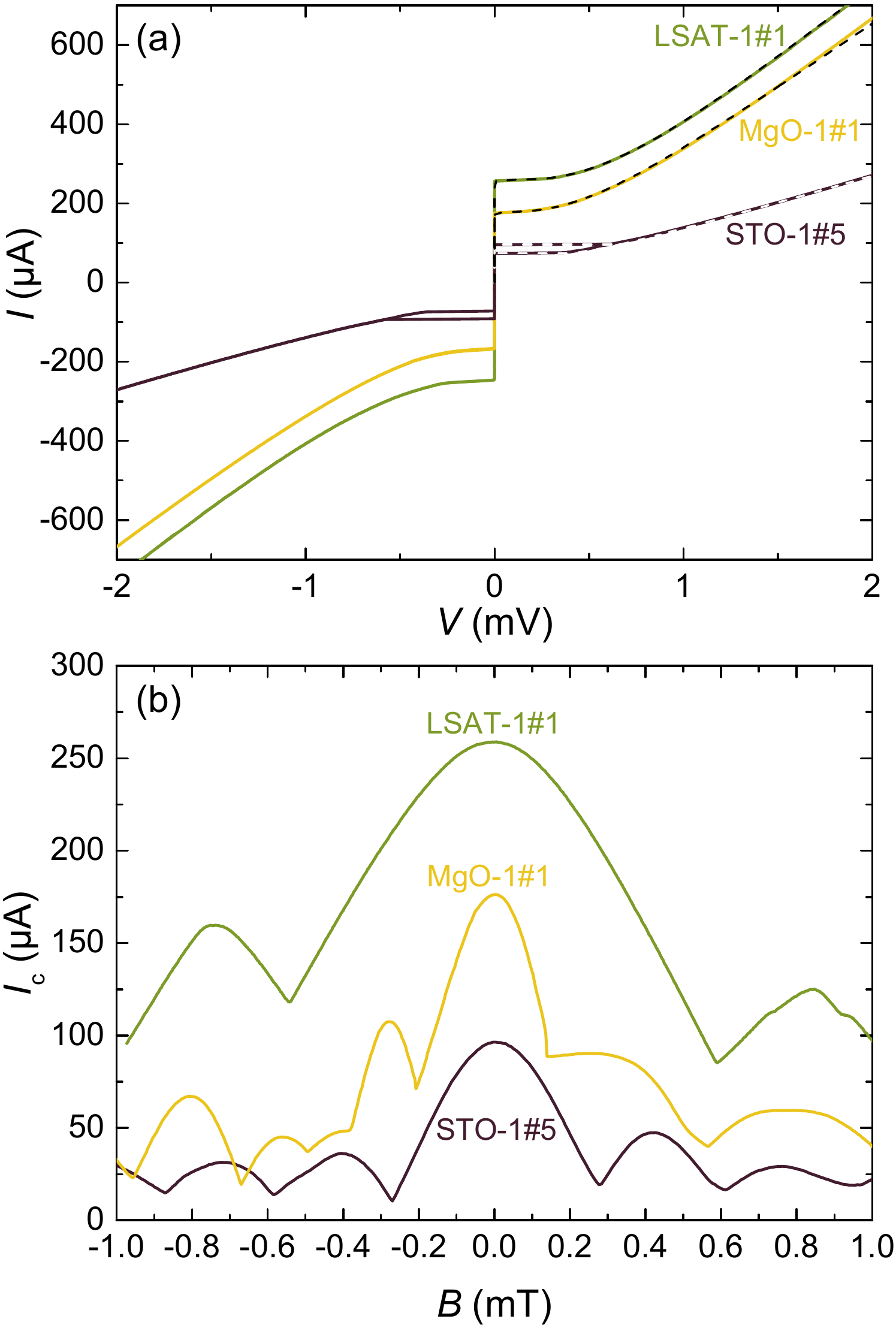}
\caption{Electric transport characteristics of He-FIB JJs fabricated on different substrates:
(a) IVCs showing experimental data (solid lines) and numerical simulation results for $I>0$ (dashed lines) and (b) $I_\mathrm{c} (B)$ patterns.}
\label{Fig:IVC-Ic(B)}
\end{figure}

In the following we present electric transport characteristics of $\sim 50$ YBCO brides that were irradiated with doses up to $D= 800 \ipnm$ to produce JJs.
We measured current $I$ vs.~voltage $V$ characteristics (IVCs) and the modulation of the critical current $I_\mathrm{c}$ in an externally applied magnetic field $B$ (perpendicular to the substrate plane) in electrically and magnetically shielded environment, with the samples at $T = 4.2 \unit{K}$ immersed in liquid He.

For all devices exhibiting IVCs that can be described by the RCSJ model, we peformed numerical simulations, including thermal noise, to determine their noise-free critical current $I_0$, normal resistance  $R_\mathrm{n}$ and capacitance $C$.
From these simulations we determined the Stewart-McCumber parameter $\beta_\mathrm{C}\equiv 2\pi I_0 R_\mathrm{n}^2 C / \Phi_0$, and also the amount of excess current $I_\mathrm{e}$, if present.

Figure~\ref{Fig:IVC-Ic(B)} shows a representative set of IVCs [Fig.~\ref{Fig:IVC-Ic(B)}(a)] and $I_\mathrm{c}(B)$ patterns [Fig.~\ref{Fig:IVC-Ic(B)}(b)] for JJs on different substrate materials.
%
%
Irradiation doses and characteristic JJ parameters are given in Table~\ref{Tab:devices}.
As shown by the simulated curves [dashed lines in Fig.~\ref{Fig:IVC-Ic(B)}(a)], the IVCs can be well described by the RCSJ model and do not show any excess current.
Only the JJ on STO shows a hysteresis in the IVC, with a JJ capacitance $C= 0.17 \unit{pF}$ and $\beta_\mathrm{C} = 2.77$ obtained from simulations.
Data regarding $\beta_\mathrm{C}$ for all JJs will be presented and discussed at the end of this section.

The $I_\mathrm{c} (B)$ patterns shown in Fig.~\ref{Fig:IVC-Ic(B)}(b) exhibit clear modulation of the critical current with applied magnetic field, however significantly deviate from a Fraunhofer-like shape expected for homogeneous JJs in the short junction limit $w\lesssim 4 \lambda_\mathrm{J}$, where
\begin{equation}
  \lambda_\mathrm{J} = \sqrt{\frac{\Phi_0}{2\pi\mu_0 d_\mathrm{eff} j_0}}
   \label{Eq:lambda_J.def}
\end{equation}
is the Josephson penetration depth with the effective JJ inductance $\mu_0 d_\mathrm{eff}$.
Our YBCO films grown on STO have a London penetration depth $\lambda_{\rm L} \approx 250 \unit{nm}$~\cite{Woelbing14,Thiel16,Martinez-Perez17,Rohner18}.
Therefore, the devices discussed here are clearly in the thin-film limit $d\ll\lambda_\mathrm{L}$, and hence the effective penetration depth $\lambda_\mathrm{eff}=\lambda_\mathrm{L}^2/d$, see Tab.~\ref{Tab:samples}, should be used to determine $d_\mathrm{eff}=2\lambda_\mathrm{eff}$ in Eq.~\eqref{Eq:lambda_J.def}.
Thus, we obtain the values of $\lambda_\mathrm{J}$ listed in Tab.~\ref{Tab:devices}, and see that all JJs are in the long JJ limit ($w> 4\lambda_\mathrm{J}$).

Moreover, since $\lambda_\mathrm{J}\ll \lambda_\mathrm{eff}$, these JJs are in the nonlocal regime~\cite{Moshe08,Abdumalikov09,Clem10a,Boris13,Kogan14a}.
In this regime the $I_\mathrm{c}(B)$ patterns can be calculated analytically~\cite{Moshe08,Kogan14a} only in the narrow JJ limit, i.e. for $w<\lambda_\mathrm{eff}$ and $w<\lambda_\mathrm{J}^\mathrm{NL}$, where $\lambda_\mathrm{J}^\mathrm{NL}=\lambda_\mathrm{J}^2/d$ is a nonlocal Josephson length~\cite{Moshe08}, see Tab.~\ref{Tab:devices}.
In our case, the above conditions are not really satisfied.
Thus, our JJs are in the intermediate regime where the exact shape of $I_\mathrm{c}(B)$ is not known.
However, we can roughly estimate the value of the penetration field $B_{\mathrm{c}1}$ (defined as a continuation of the first lobe of the $I_{\rm c}(B)$ dependence down to $I_\mathrm{c}=0$) using (i) the local long JJ model and (ii) the narrow non-local JJ model~\cite{Moshe08} and compare them with experimentally measured values of $B_{\mathrm{c}1}^\mathrm{ex}$, given in Tab.~\ref{Tab:devices}.
The local long JJ model yields $B_{\mathrm{c}1}^\mathrm{LO}=\Phi_0/(\pi d_\mathrm{eff} \lambda_J)$, whereas for a non-local narrow JJ~\cite{Moshe08} $B_{\mathrm{c}1}^\mathrm{NL}=\Phi_0/(0.715w^2)$ (independent of $j_0$).
By comparing these values with the experimental ones (see Tab.~\ref{Tab:devices}), we see that $B_{\mathrm{c}1}^\mathrm{ex}$ for STO-1\#5 and MgO-1\#1 are better described by the non-local theory, while $B_{\mathrm{c}1}^\mathrm{ex}$ for LSAT-1\#1 is closer to a local long JJ.

A detailed study of the $I_\mathrm{c}(B)$ patterns of our He-FIB induced JJs is out of the scope of the work presented here.
Typically, our JJs on MgO have more irregular $I_\mathrm{c}(B)$ patterns than those on STO and LSAT [cf.~Fig.~\ref{Fig:IVC-Ic(B)}(b)].
This indicates more inhomogeneous barrier properties of He-FIB induced JJs on MgO, and may either be attributed to the much larger lattice mismatch between MgO and YBCO and correspondingly poorer crystalline quality of YBCO films or stronger charging of the MgO substrates which was observed during the He-FIB process.

In the following, we analyze the scaling of characteristic JJ properties $j_\mathrm{c}$, $\rho_\mathrm{n}$ and $j_\mathrm{c}\rho_\mathrm{n}$ with irradiation dose $D$.
Note that devices irradiated with the lowest doses do not show JJ behavior.
Therefore we denote here all critical currrent densities as $j_\mathrm{c}$.
For all devices showing RCSJ behavior, however, the values given here refer to the noise-free values of $j_0$ obtained from numerical simulations.

\begin{figure}[b]
\includegraphics[width=\columnwidth]{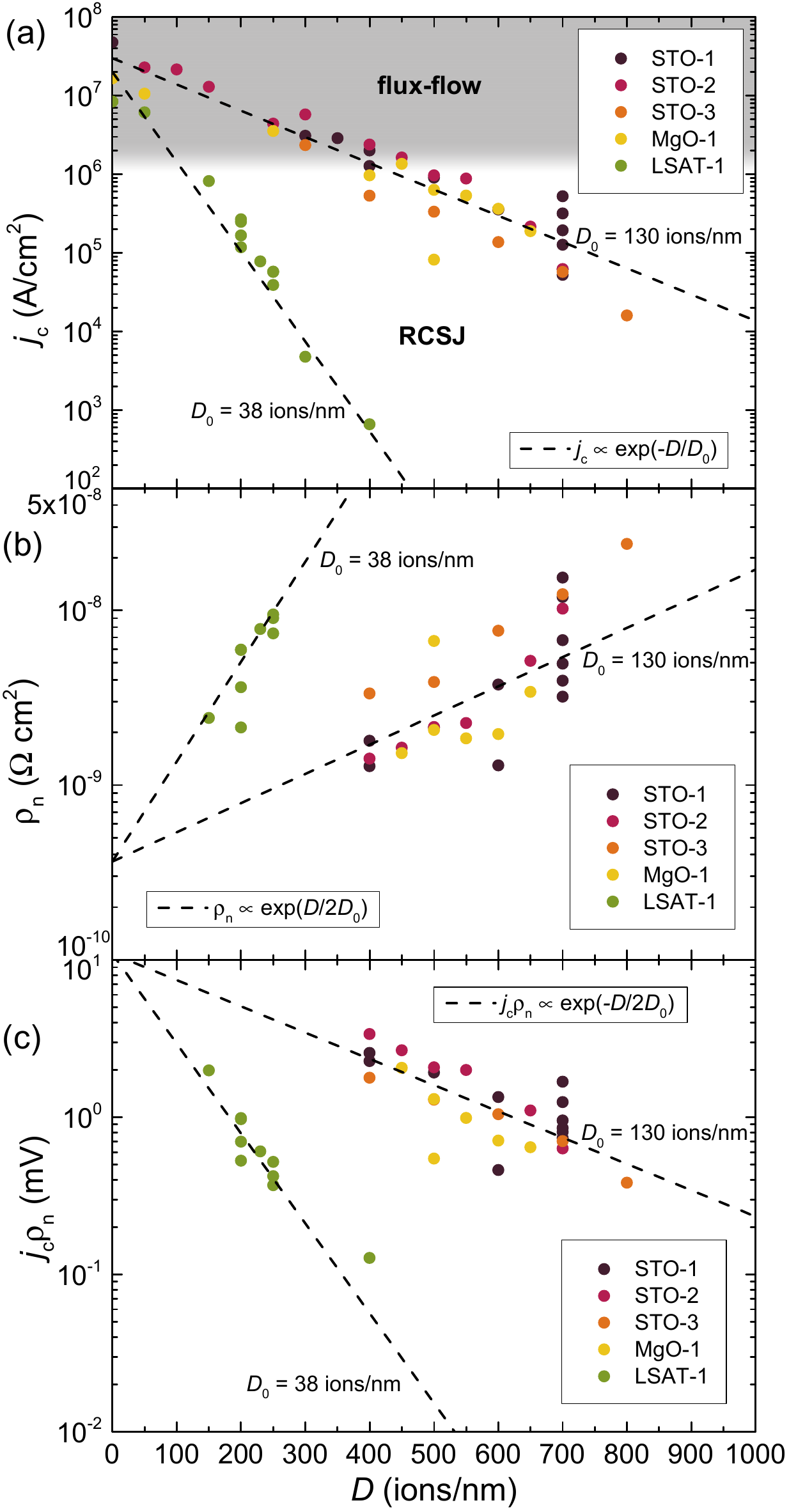}
\caption{Dependence of device parameters on He ion line dose $D$ for various samples on different substrates:
(a) $j_\mathrm{c}(D)$, (b) $\rho_\mathrm{n}(D)$ and (c) $j_\mathrm{c}\rho_\mathrm{n}(D)$.
The dashed lines indicate the approximate scaling behavior as discussed in the text.
}
\label{Fig:scaling(D)}
\end{figure}

Figure~\ref{Fig:scaling(D)}(a) summarizes $j_\mathrm{c} (D)$ for all investigated devices.
We attribute the significant scatter (cf., e.g., the data points for $700 \ipnm$ on STO-1) to instabilities in the fabrication process which we have not yet optimized.
For instance, slight variations in He-FIB focus spot size or beam current will both affect the barrier properties.
In spite of this scatter, we clearly find an exponential decay $j_\mathrm{c}(D)\approx j_{\mathrm{c},0}\exp(-D/D_0)$, with $j_{\mathrm{c},0}= 3\times 10^7 \unit{A/cm}^2$ and with $D_0=38 \ipnm$ for LSAT-1 and $D_0=130 \ipnm$ for the other chips.
The reason for the much stronger decay of $j_\mathrm{c} (D)$ on LSAT compared to the devices on STO or MgO has not been clarified yet.
Clearly, we do not find a correlation of $D_0$ with YBCO film thickness or crystalline quality (cf.~FWHM values in table~\ref{Tab:samples}).

For $j_\mathrm{c}\lesssim 2 \unit{MA/cm^2}$, the IVCs show RCSJ-like behavior, whereas devices with higher critical current densities yield flux-flow type IVCs, as indicated by the grey area in Figure~\ref{Fig:scaling(D)}(a).
Altoghether, the range of variation of $j_\mathrm{c}$ covers five orders of magnitude.
We note that an exponentially decaying behavior of $j_\mathrm{c}$ is well known from cuprate grain boundaries, where $j_\mathrm{c}$ decays exponentially with the grain boundary misorientation angle $\Theta$~\cite{Gross91a,Hilgenkamp02,Graser10}.
In a theoretical analysis of cuprate grain boundary JJs, Graser {\it et al.}~\cite{Graser10} related the exponential decay of $j_\mathrm{c}(\Theta)$ to charging of the interface near defects induced by the structural distortions at the grain boundary.
For the He-FIB-induced JJ barriers, the locally induced defect structure is not known yet, and it remains to be clarified whether a similar charging mechanism is responsible for the exponential decay of $j_\mathrm{c} (D)$.
A simple explanation of the exponential decay of $j_\mathrm{c}$ with increasing $D$ would be a linear increase of the JJ barrier thickness.
As stated in Sec.~\ref{subsec:TEM}, the STEM analysis yields a roughly linear increase of the amorphous track width with increasing $D$ for doses above $1000 \ipnm$.
However, for lower doses the STEM data do not allow us to make a statement on the width of the induced defect regions and their scaling with $D$.

\begin{figure}[t]
\includegraphics[width=\columnwidth]{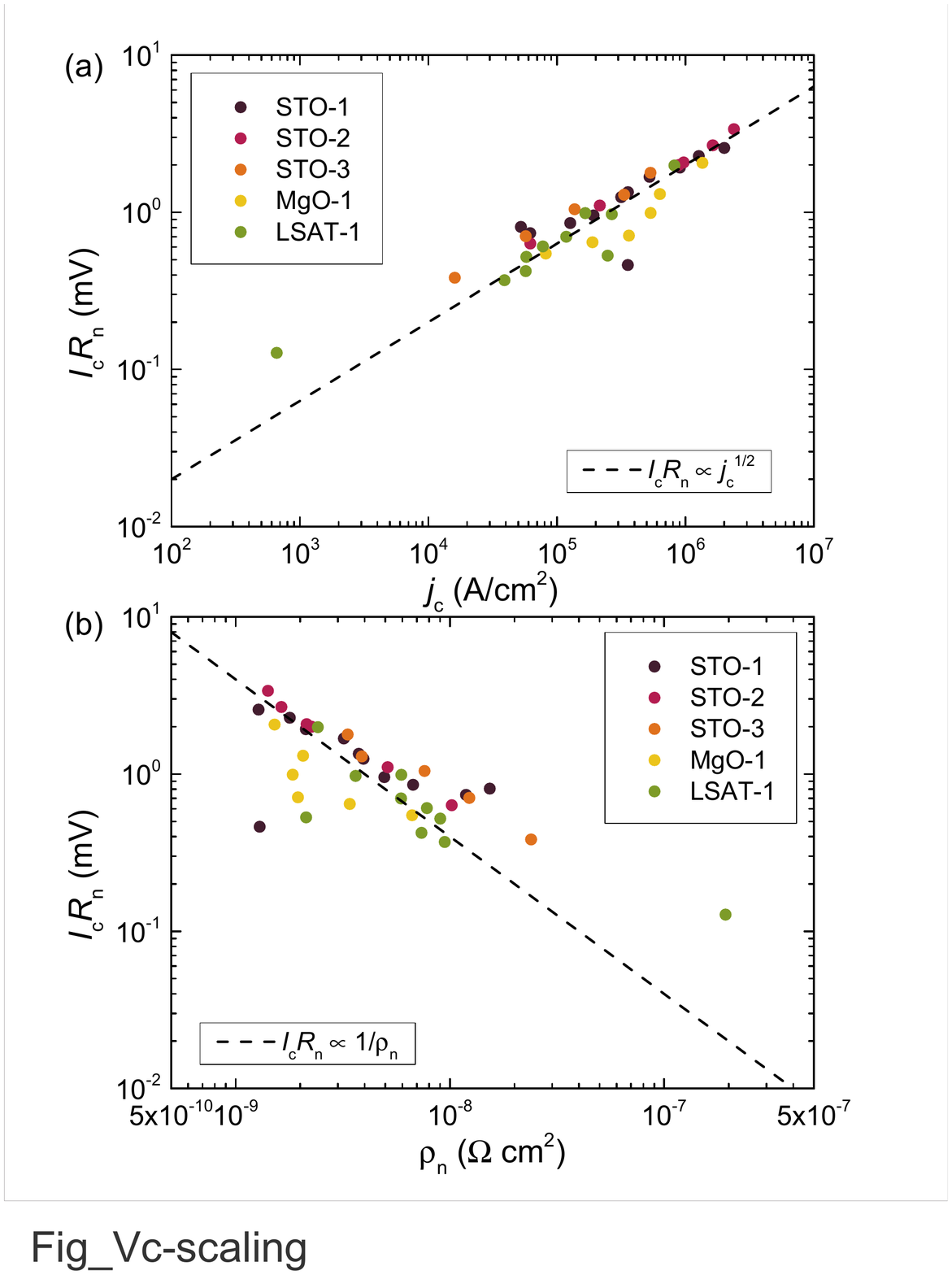}
\caption{Characteristic voltage $I_\mathrm{c}R_\mathrm{n}$ for He-FIB JJs with RCSJ-like IVCs on different substrates (a) vs.~critical current density $j_\mathrm{c}$ and (b) vs.~resistance times area $\rho_\mathrm{n}$.
Dashed lines indicate $I_\mathrm{c}R_\mathrm{n}\propto\sqrt{j_\mathrm{c}}$ in (a) and $I_\mathrm{c}R_\mathrm{n}\propto 1/\rho_\mathrm{n}$ in (b) as discussed in the text.}
\label{Fig:Vc-scaling}
\end{figure}

Our analysis of the IVCs of He-FIB JJs produced with variable dose also yields a systematic scaling of the resistance times area product $\rho_\mathrm{n}\equiv R_\mathrm{n}wd\approx\rho_{\mathrm{n},0}\exp(D/2D_0)$ with $\rho_{\mathrm{n},0}=0.37 \unit{n \Omega \, cm^2}$, i.e., $\rho_\mathrm{n}$ increases exponentially with $D$ as shown in Fig.~\ref{Fig:scaling(D)}(b).
Interestingly, the stronger decay of $j_\mathrm{c}(D)$ for JJs on LSAT comes along with a correspondingly stronger increase in $\rho_\mathrm{n}(D)$, i.e., we can use the same values of $D_0$ for the scaling of $\rho_\mathrm{n}(D)$ as used for the scaling of $j_\mathrm{c}(D)$.
Accordingly, the characteristic voltage $V_\mathrm{c}=j_\mathrm{c}\rho_\mathrm{n}$ also shows an exponential scaling $V_\mathrm{c}\approx V_{\mathrm{c},0}\exp(-D/2D_0)$ with $V_{\mathrm{c},0}=j_{\mathrm{c},0}\,\rho_{\mathrm{n},0}=11 \unit{mV}$, as shown in Fig.~\ref{Fig:scaling(D)}(c).
We note here, that Figs.~\ref{Fig:scaling(D)}(b),(c) contain only data points that correspond to RCSJ-type IVCs.

The fact that our analysis of the scaling of characteristic JJ properties ($j_\mathrm{c}$, $\rho_\mathrm{n}$ and $V_\mathrm{c}$) with $D$ can be described by the same values of $D_0$ indicates a universal scaling of $V_\mathrm{c}$ with either $j_\mathrm{c}$ or $\rho_\mathrm{n}$ independent of substrate material.
This is shown in Fig.~\ref{Fig:Vc-scaling}, where we display $V_\mathrm{c}(j_\mathrm{c})$ and $V_\mathrm{c}(\rho_\mathrm{n})$ for all investigated JJs.
Despite the significant scatter in the data, a clear trend is visible, that can be described by $I_\mathrm{c}R_\mathrm{n}\approx V_{\mathrm{c},1}(j_\mathrm{c}/j_{\mathrm{c},1})^{1/2}$ (dashed line in Fig.~\ref{Fig:Vc-scaling}(a)), with $V_{\mathrm{c},1}=2 \unit{mV}$ and $j_{\mathrm{c},1}=10^6 \unit{A/cm^2}$ and by $I_\mathrm{c}R_\mathrm{n}\approx V_{\mathrm{c},1}(\rho_{\mathrm{n},1}/\rho_\mathrm{n})$ (dashed line in Fig.~\ref{Fig:Vc-scaling}(b)) with $\rho_{\mathrm{n},1}=2 \unit{n \Omega\, cm^2}$.
We note that an approximate scaling $I_\mathrm{c}R_\mathrm{n}\propto\sqrt{j_\mathrm{c}}$ and $I_\mathrm{c}R_\mathrm{n}\propto 1/\rho_\mathrm{n}$ has also been observed for many cuprate grain boundary JJs and other JJ types in cuprate superconductors, albeit with a slightly larger $V_{\mathrm{c},1}$ for the same $j_{\mathrm{c},1}$ and $\rho_{\mathrm{c},1}$~\cite{Gross97}.
However, we should also note that the existence or absence of a universal scaling of $I_{\rm c}R_{\rm n}$ versus $j_{\rm c}$ or $\rho_{\rm n}$ for all cuprate JJs and its origin has been discussed controversially in the literature; see e.g.~Refs.~\onlinecite{Gross97,Pauza97,Hilgenkamp02}.
At least, for oxygen-depleted grain boundaries \cite{Hilgenkamp02} there seems to be consensus on the same scaling as we see in our He-FIB JJs.
This is probably not surprising, because the He-FIB irradiation induces such an oxygen depletion\cite{Cybart15}.

So, obviously, for achieving large values of $V_\mathrm{c}$ one should use doses that are as small as possible to obtain large values of $j_\mathrm{c}$, but still provide JJs with RCSJ-type IVCs.
Moreover, for fabricating SQUIDs (cf.~Sec.~\ref{sec:SQUIDs}) one wants to have non-hysteretic IVCs, i.e.~$\beta_\mathrm{C}\lesssim 1$.
To address these issues, we determined from RCSJ simulations the dependencies of $\beta_\mathrm{C}$ and of the excess current densities $j_\mathrm{e}$ of our He-FIB JJs on $j_\mathrm{c}$.

\begin{figure}[t]
\includegraphics[width=\columnwidth]{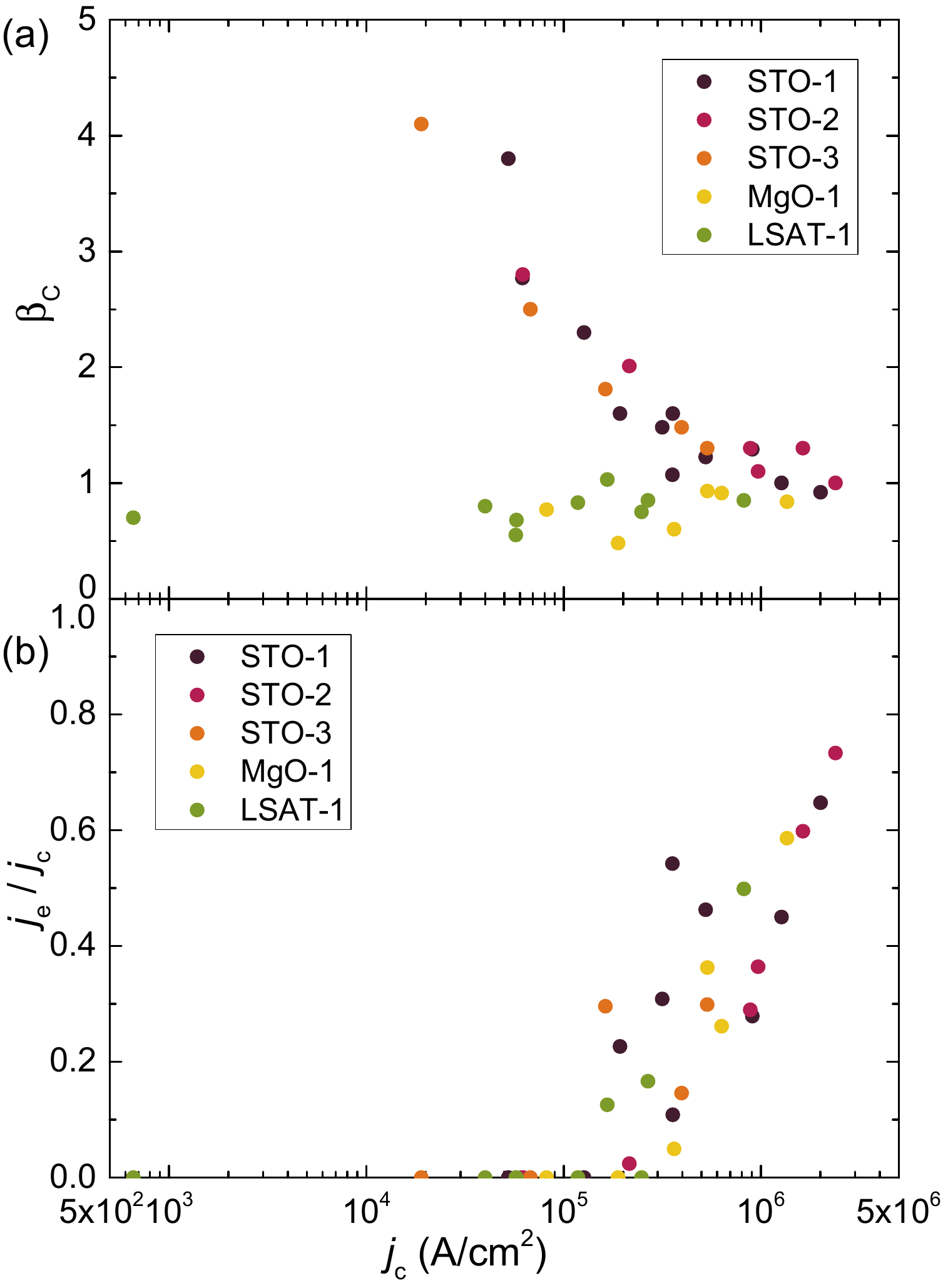}
\caption{(a) Stewart-McCumber parameter $\beta_\mathrm{C}(j_\mathrm{c})$ and  (b) normalized excess current $j_\mathrm{e}/j_\mathrm{c}(j_\mathrm{c})$ for He-FIB JJs on different substrates.}
\label{Fig:betaC-je}
\end{figure}

Figure \ref{Fig:betaC-je}(a) shows $\beta_\mathrm{C}(j_\mathrm{c})$.
We clearly find a significant difference for devices on STO as compared to those on LSAT or MgO.
While for devices on LSAT and MgO the values of $\beta_\mathrm{C}$ are essentially independent of $j_\mathrm{c}$ and yield values between $\sim 0.5$ and 1, for devices on STO $\beta_\mathrm{C}$ is always above 1 (up to $\sim 4$) and decays with increasing $j_\mathrm{c}$.
This reflects the fact that we observed hysteretic IVCs at $4.2 \unit{K}$ for all JJs on STO substrates and non-hysteretic IVCs for all JJs on MgO and LSAT.
We attribute this behavior to a significant stray capacitance contribution caused by the large permittivity of STO at low $T$~\cite{Beck95}.
However, a more detailed analysis of this behavior would require a systematic variation of the JJ width, which we have not performed for the present study.
For YBCO grain boundary JJs, the capacitance $C$ per area $A$ has been found to vary roughly \cite{Hilgenkamp02} within $10^{-6}$ to $10^{-4}$\,F/cm$^2$, with a significant increase with increasing $j_{\rm c}$ from roughly $10^3$ to $10^6$\,A/cm$^2$.
In contrast, for our He-FIB JJs we find for most devices $C/A$ to scatter within the range from $5\times 10^{-5}$ to $2\times 10^{-4}$\,F/cm$^2$ with no clear dependence on $j_{\rm c}$ in the range from $2\times 10^4$ to $2\times 10^6$\,A/cm$^2$ (not shown).

Figure \ref{Fig:betaC-je}(b) shows the normalized excess current density $j_\mathrm{e}/j_\mathrm{c}(j_\mathrm{c})$.
For the largest values of $j_\mathrm{c}$, we do find excess currents that decay with decreasing $j_\mathrm{c}$ and finally disappear at $j_\mathrm{c}\approx 2\times 10^5 \unit{A/cm^2}$.
This behavior seems to be independent of the substrate material.
The appearance of excess currents has also been reported for cuprate grain boundary JJs and electron-beam damaged JJs with large current densities \cite{Pauza97,Hilgenkamp02}.
Such JJs are often modeled as superconductor/normal conductor/superconductor (SNS) JJs.
For He-FIB JJs in YBCO films, Cybart {\it et al.} \cite{Cybart15} report on the transition from SNS-type to SIS-type JJs (I: insulator) upon increasing the irradiation dose. Our results are consistent with this observation; a more detailed analysis, however, requires transport measurements at variable temperature, which we have not performed so far in detail.

To conclude this section, we can state that for obtaining devices that do not exhibit excess currents, one should not exceed $j_\mathrm{c}\approx 10^5 \unit{A/cm^2}$.
For devices on STO, however, such low $j_\mathrm{c}$ values come with values of $\beta_\mathrm{C}$ clearly above 1, i.e.~with hysteretic IVCs.
The $I_\mathrm{c}(B)$ patterns of devices on MgO, on the other hand, show the strongest deviations from a Fraunhofer-like behavior.
Hence, at the current state it seems that, among the substrate materials investigated here, He-FIB JJ devices on LSAT are most promising for applications.

\section{He-FIB-induced dc SQUIDs}
\label{sec:SQUIDs}

By irradiation with high He ion doses (typically for $D\gtrsim 1000 \ipnm$), $j_\mathrm{c}$ can be fully suppressed, as shown in Fig.~\ref{Fig:R(T)}, and the He-FIB-induced barriers can be made highly resistive (reaching even $\mathrm{G \Omega}$ resistances at $4.2 \unit{K}$ for $D \gtrsim 5000 \ipnm$).
This offers the possibility to define the sample geometry -- ultimately on the nm-scale -- via direct-write lithography, without removing material by milling~\cite{Cho18,Cho18a}.

\begin{figure}[t!]
\includegraphics[width=\columnwidth]{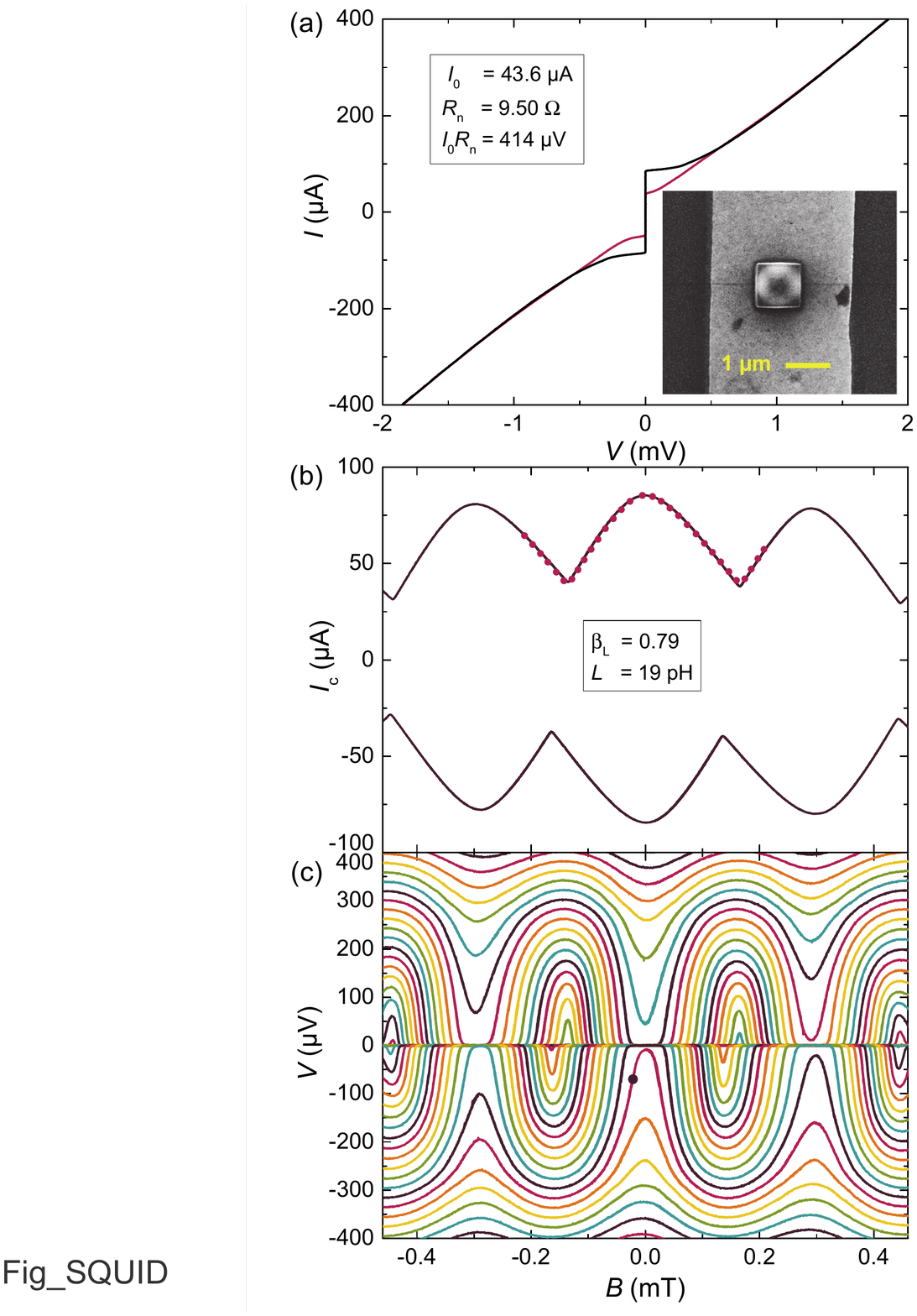}
\caption{Electric transport characteristics of a He-FIB dc SQUID fabricated on LSAT-1\#2.
%
%
(a) IVCs for applied flux yielding maximum (black) and minimum (red) positive critical current.
The inset shows an SEM image of a similar SQUID fabricated on STO-1.
%
%
(b) Critical current vs.~magnetic field: experimental data (solid lines) and numerical simulation (symbols).
(c) Voltage-flux dependence for bias currents $I_\mathrm{b}$ within $\approx\pm 110 \unit{\mu A}$ in steps of $4 \unit{\mu A}$.
Dark dot near center indicates the working point for noise measurement.
}
\label{Fig:SQUIDtransp}
\end{figure}

The combination of He-FIB irradiation with medium and high doses provides a simple way of fabricating dc SQUIDs from photolithographically prepatterned YBCO thin film bridges on single crystal substrates, with tailored JJ properties and SQUID inductance.
We used this approach to fabricate simple micro- and nanoSQUIDs on STO, MgO and LSAT substrates by first scanning over a square shaped area at the center of the prepatterned YBCO bridge to define the SQUID `hole' (i.e.~a highly resistive, magnetically transparent area) and a subsequent linescan across the whole width of the bridge using a medium dose to produce the JJs.
An SEM image of such a SQUID is shown in the inset of Fig~\ref{Fig:SQUIDtransp}(a) for a device fabricated on STO-1 with a $1 \times 1 \unit{\mu m^2}$ hole (irradiated with $D_\mathrm{A}=4000 \ipsqnm$).
Again, the location of the JJs (irradiated with $D=700 \ipnm$) and the SQUID hole is visible via the He-FIB induced carbon deposition.

Since the hysteresis in the IVCs severely limits the performance of SQUIDs on STO substrates, the electric transport and noise data shown in the following were measured on a device fabricated on LSAT-1.
On the LSAT substrate, however, SEM imaging was only possible in poor quality due to charging of the substrate.
For the device on LSAT, the SQUID hole was defined as a $300 \times 300 \unit{nm^2}$ sized square, irradiated with $D_\mathrm{A}=4000 \ipsqnm$.
The JJs were fabricated by a linescan with $D=230 \ipnm$ and have a width of $\sim 2 \unit{\mu m}$ each.

\begin{figure}[t]
\includegraphics[width=\columnwidth]{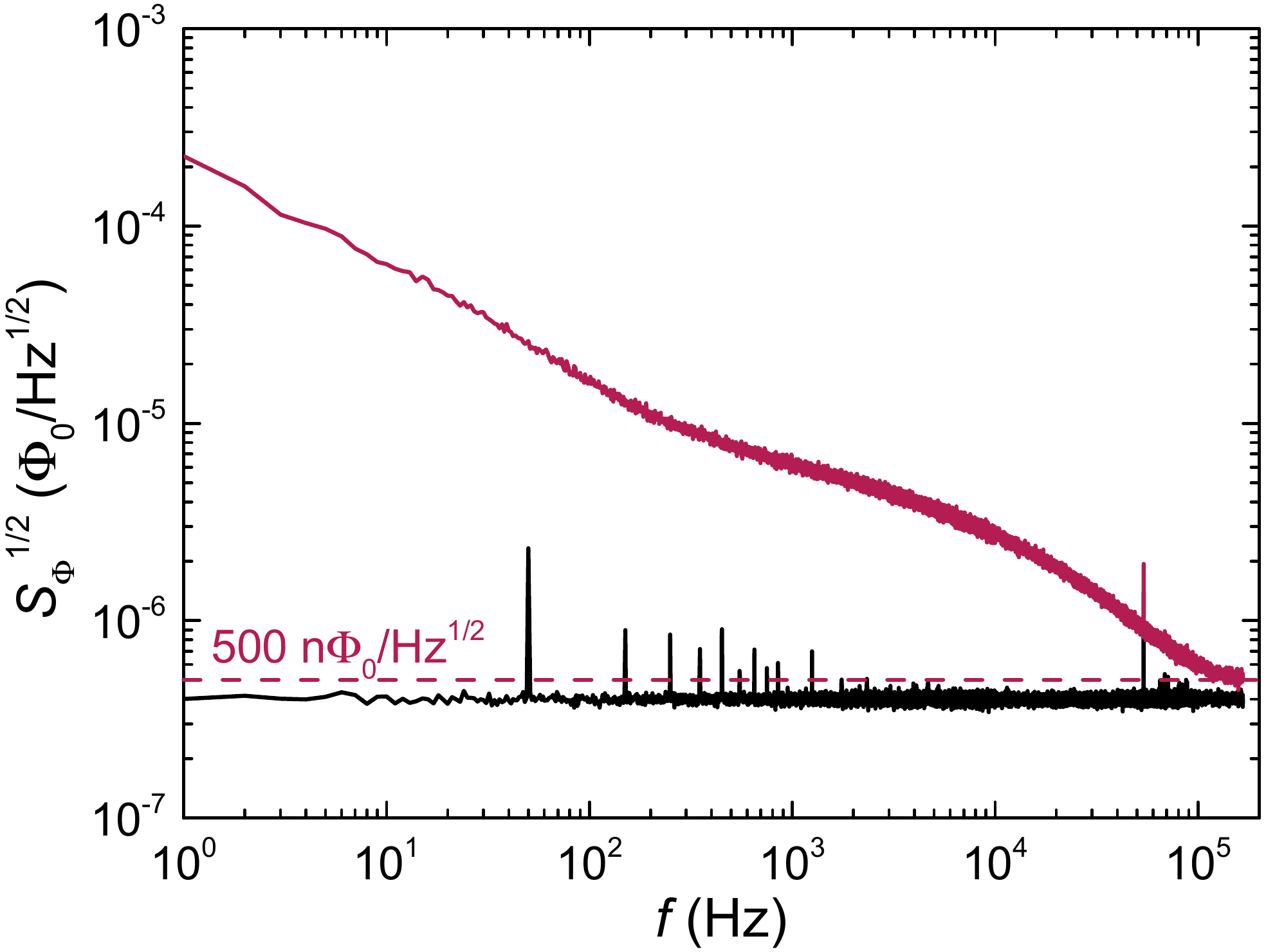}
\caption{Noise characteristics of a He-FIB dc SQUID fabricated on LSAT-1\#2.
%
%
Flux noise spectra of the SQUID (red) and electronics background (black).
Dashed line indicates upper limit for thermal white noise $S_{\Phi,\mathrm{w}}^{1/2} = 500 \unit{n\Phi_0/Hz^{1/2}}$ of the SQUID.
}
\label{Fig:SQUIDnoise}
\end{figure}

Figure~\ref{Fig:SQUIDtransp}(a) shows the IVCs for different applied magnetic flux to yield maximum (black) and minimum (red) positive critical current, exhibiting neither hysteresis nor excess current.
From RCSJ simulations we determine a mean per-junction critical current $I_0 = 43.6 \unit{\mu A}$ and normal state resistance $R_\mathrm{n} = 9.50 \unit{\Omega}$ ($V_\mathrm{c} = 414 \unit{\mu V}$) and $\beta_\mathrm{C}=0.74$.
%
The dependence of the critical current $I_\mathrm{c,s}$ of the SQUID on magnetic field is shown in Fig.~\ref{Fig:SQUIDtransp}(b) as solid lines, together with the numerically simulated dependence, based on the RCSJ model~\cite{Chesca-SHB-2} (symbols).
From the simulations we extract the screening parameter $\beta_L\equiv 2I_\mathrm{c}L/\Phi_0 =$ 0.79, inductance $L = 19 \unit{pH}$, and asymmetry parameters \cite{Chesca-SHB-2} for the critical current $\alpha_I =0.145$ and inductance $\alpha_L = -0.15$.
%
Figure \ref{Fig:SQUIDtransp}(c) shows the voltage-flux dependence of the SQUID for a range of bias currents $I_\mathrm{b} \approx \pm 110 \unit{\mu A}$ in steps of $4 \unit{\mu A}$.
The dark dot near the center on the curve for $I_\mathrm{b} = -85 \unit{\mu A}$ indicates the working point with a transfer function $V_\Phi = 2.1 \unit{mV/\Phi_0}$ which was used for the noise measurement shown in Fig.~\ref{Fig:SQUIDnoise}.

To determine the spectral density of flux noise $S_\Phi$ vs. frequency $f$ of the SQUID we used a Magnicon SEL-1 SQUID electronics~\cite{Magnicon-SEL-1} in direct readout mode~\cite{Drung03}.
Figure~\ref{Fig:SQUIDnoise} shows the measured root-mean-square (rms) spectral density of flux noise $S_\Phi^{1/2}(f)$ of the SQUID (red), together with the background noise of the readout electronics (black).
The noise spectrum is dominated by frequency-dependent excess noise, scaling roughly as $S_\Phi\propto 1/f$ ($1/f$ noise), with a small bump (at $\sim 10 \unit{kHz}$).
This excess noise extends all the way up to the cut-off frequency $f_\mathrm{el} = 166 \unit{kHz}$ of our readout electronics (limited by the sampling rate of the ADC), where it almost reaches the noise floor of the readout electronics.
For YBCO SQUIDs one typically finds a strong $1/f$ noise contribution due to $I_\mathrm{c}$ fluctuations in the JJs~\cite{Koelle99}.
Probably, this is also the case for the SQUID presented here.
To further clarify this, one should operate the SQUIDs in current bias reversal mode \cite{Drung-SHB-4} which eliminates the excess noise contribution from $I_\mathrm{c}$ fluctuations.
We have already successfully demonstrated this approach for YBCO nanoSQUIDs based on grain boundary JJs~\cite{Schwarz15}, however, for the simple SQUID layout without a suitable on-chip flux coupling structure, as used in this work, it was not possible to use this approach.
At least, from the noise data shown in Fig.~\ref{Fig:SQUIDnoise}, we can give an upper limit for the thermal white noise $S_{\Phi,\mathrm{w}}^{1/2} \lesssim 500 \unit{n \Phi_0/Hz^{1/2}}$, which is impressively low for a $L\approx 20\,$pH SQUID.

\section{Conclusions}
\label{sec:Conclusions}

We have successfully demonstrated the fabrication of YBCO Josephson junctions and dc SQUIDs by using a focused He ion beam which locally modifies epitaxially grown YBCO thin films and allows us to `write' Josephson barriers and insulating areas with high spatial resolution. 
The analysis of the electric transport properties at $4.2 \unit{K}$ of our He-FIB-induced structures confirm and extend earlier results obtained by Cybart and co-workers~\cite{Cybart15,Cho15,Cho18}.

We studied in detail the dependence of characteristic JJ properties on irradiation dose for devices on STO, MgO and LSAT substrates.
Upon increasing the irradiation dose, we find a transition from flux-flow to RCSJ-like behavior with some excess current contribution that vanishes upon further increasing the dose.
Moreover, we find an exponential decay of the critical current density $j_0$ with increasing dose.
For currently unclear reasons, this decay is much faster for devices on LSAT as compared to devices on STO and MgO.
Another major difference regarding JJ behavior on different substrates is the observation of hysteretic IVCs for devices on STO, while devices on LSAT and MgO show no hysteresis.
We attribute the hysteresis in the IVCs to a stray capacitance contribution from the STO substrates.
The analysis of the characteristic voltage $V_\mathrm{c}$ of the fabricated JJs yields an approximate scaling $V_\mathrm{c}\propto\sqrt{j_0}$.

Altogether, He-FIB JJs offer new perspectives for creating Josephson devices, due to the possibility to control the JJ properties by irradiation dose even on the same substrate and to place the JJs at virtually arbitrary positions.
This flexibility obviously offers an enormous advantage for creating advanced devices, in particular employing multi-JJ configurations.
Our detailed analysis of the JJ properties can be very helpful for designing optimized devices for applications.

Moreover, irraditiation with high dose drives the material highly resistive.
In this regime, our STEM analysis shows the creation of amorphous tracks in the YBCO films, which for not too high doses still have a lateral extension down to only a few nm.
This observation indicates that He-FIB irradiation is a promising tool for nanopatterning (without removal of material) of YBCO films with ultra-high resolution.
We used this feature to produce dc SQUIDs by patterning both the JJs and the SQUID loop by He-FIB irradiation.
For a device on LSAT we demonstrate very low flux noise in the thermal white noise regime.
The observed low-frequency excess noise still has to be investigated in detail in further studies.
Although we have not yet pushed  to the ultimate limit of miniaturization, we envisage that He-FIB irradiation should be ideally suited for the realization of ultra-low noise nanoSQUIDs~\cite{Granata16,Martinez-Perez17a} due to the high spatial resolution of helium ion microscopy.

\acknowledgments

B.~ M\"{u}ller acknowledges funding by the German Academic Scholarship Foundation.
We gratefully acknowledge fruitful discussions with S.~Cybart, V.~G.~Kogan, R.~G.~Mints and R.~Menditto and technical support by M.~Turad and R.~L\"offler (LISA$^+$), W.~Nisch and C.~Warres (NMI) and by C.~Back.
This work was supported by the  COST action NANOCOHYBRI (CA16218).

\bibliography{YBCO-HIM-JJs_v12b}

\end{document}